\documentstyle[sprocl,epsf]{article}

\def\be{\begin{equation}}
\def\ee{\end{equation}}
\def\bea{\begin{eqnarray}}
\def\eea{\end{eqnarray}}
\newcommand{\rmi}[1]{{\mbox{\scriptsize #1}}}

\begin{document}

\title{STATIC MASS SCALES IN HOT GAUGE THEORIES}

\author{O. PHILIPSEN}

\address{CERN Theory Division, 1211 Geneva 23, Switzerland}


\maketitle\abstracts{ 
The static sectors of the electroweak Standard Model and QCD at finite 
temperatures are described by
3d SU(N) Higgs models with scalars in the fundamental and adjoint representation, 
respectively.
I summarize the non-perturbative physics of these theories 
like mass spectrum, string tension, screening of the static potential
and non-perturbative corrections to the Debye mass,
as obtained from recent lattice simulations.
It is observed that in the 3d theory corresponding to the
purely magnetic sector there is a
larger hierarchy between different quantities than between 
the electric ($\sim gT$) and magnetic ($\sim g^2T$) sectors of 
finite T gauge theories, which are 
not generally well separated.
}

\section{Introduction}

Gauge theories in 3d play an important role
for high temperature particle physics, since they
constitute the Matsubara zero mode sector of finite temperature quantum field
theories in the imaginary time formalism. In the framework of
dimensional reduction they emerge as 
effective theories describing all static properties and the equilibrium
thermodynamics of the original 4d finite T theory.
Hence, any static physical quantity
of a finite T theory
must have a corresponding quantity in the effective 3d theory, with
which it agrees up to a (perturbative) part
due to the non-zero Matsubara modes. For example, a static screening length 
in finite T field theory
is defined as the exponential decay of some {\it spatial} correlation function.
In the 3d theory,
the direction of the correlation may be taken to be (euclidean) time,
and hence the same quantity appears in the spectrum or some other
physical property of the (2+1)d theory.
While dimensional reduction is a perturbative procedure, 
the resulting effective theories for 
the symmetric phase of the electroweak Standard Model and hot QCD are
3d SU(N) gauge + fundamental and adjoint Higgs models,
respectively, in their confining phases and hence entirely 
non-perturbative.

Here I discuss the 
physical properties of the 3d SU(2) Higgs models with fundamental and
adjoint scalar fields. The scale is set by the dimensionful gauge coupling 
$g_3^2=g^2T$.
Further, the physics of both models is fixed by 
two parameters $x$ and $y$, which
are dimensionless ratios 
of the scalar coupling and bare mass with the gauge 
coupling, respectively. For
effective high T theories, $x,y$ are fixed by dimensional reduction \cite{dimr}
as functions of T and the Higgs mass (fundamental) or of T alone (adjoint).
Both models have confinement and Higgs regions in their phase diagram,
which are separated by a first order phase transition for small $x$, but 
analytically connected \cite{pt} for large $x$.

\section{Mass spectrum}

\begin{figure}[t]
\vspace*{-0.7cm}
\begin{center}
\leavevmode
\epsfysize=3.6cm
\epsfbox{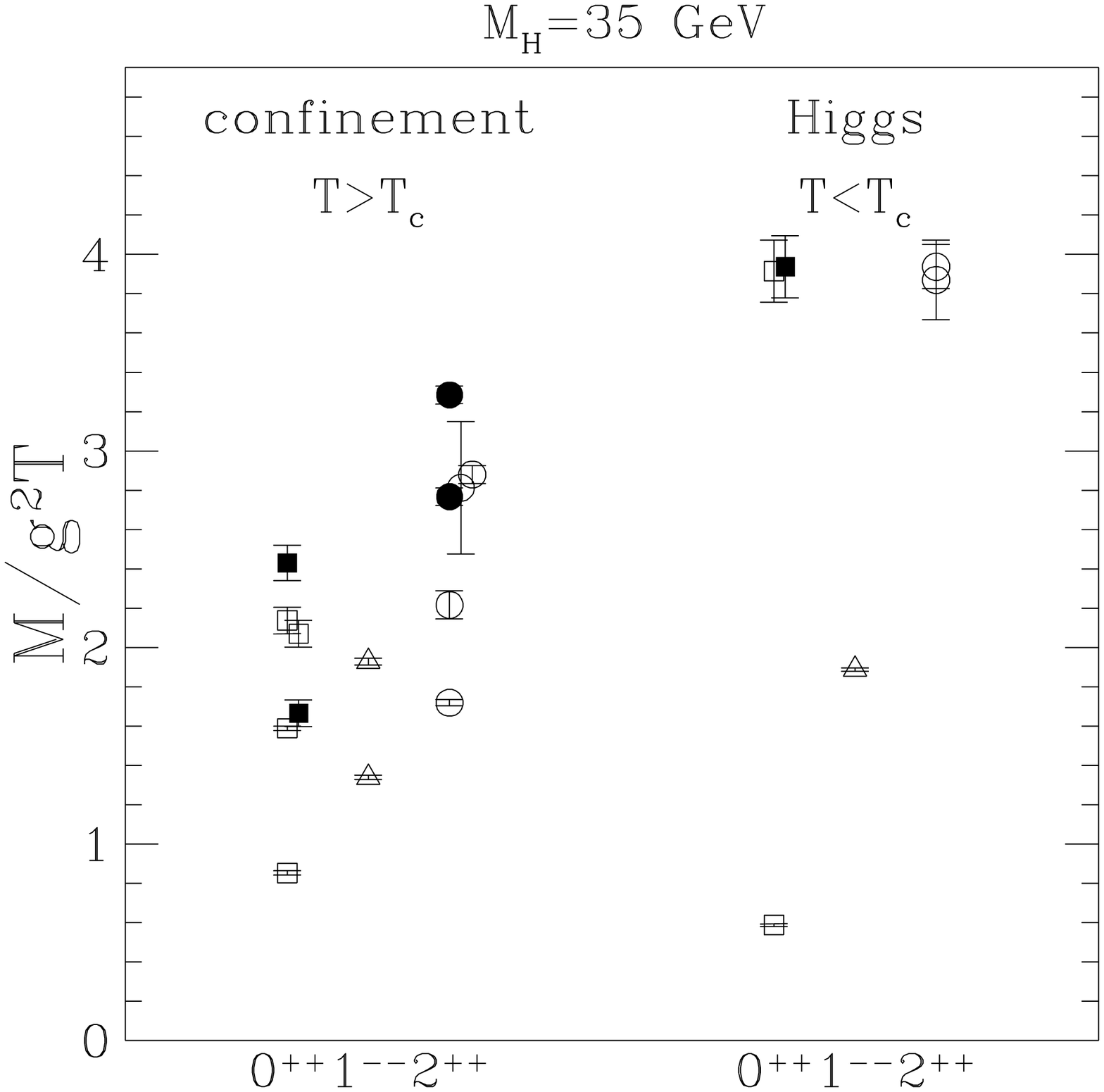}
\leavevmode
\epsfysize=3.6cm
\epsfbox{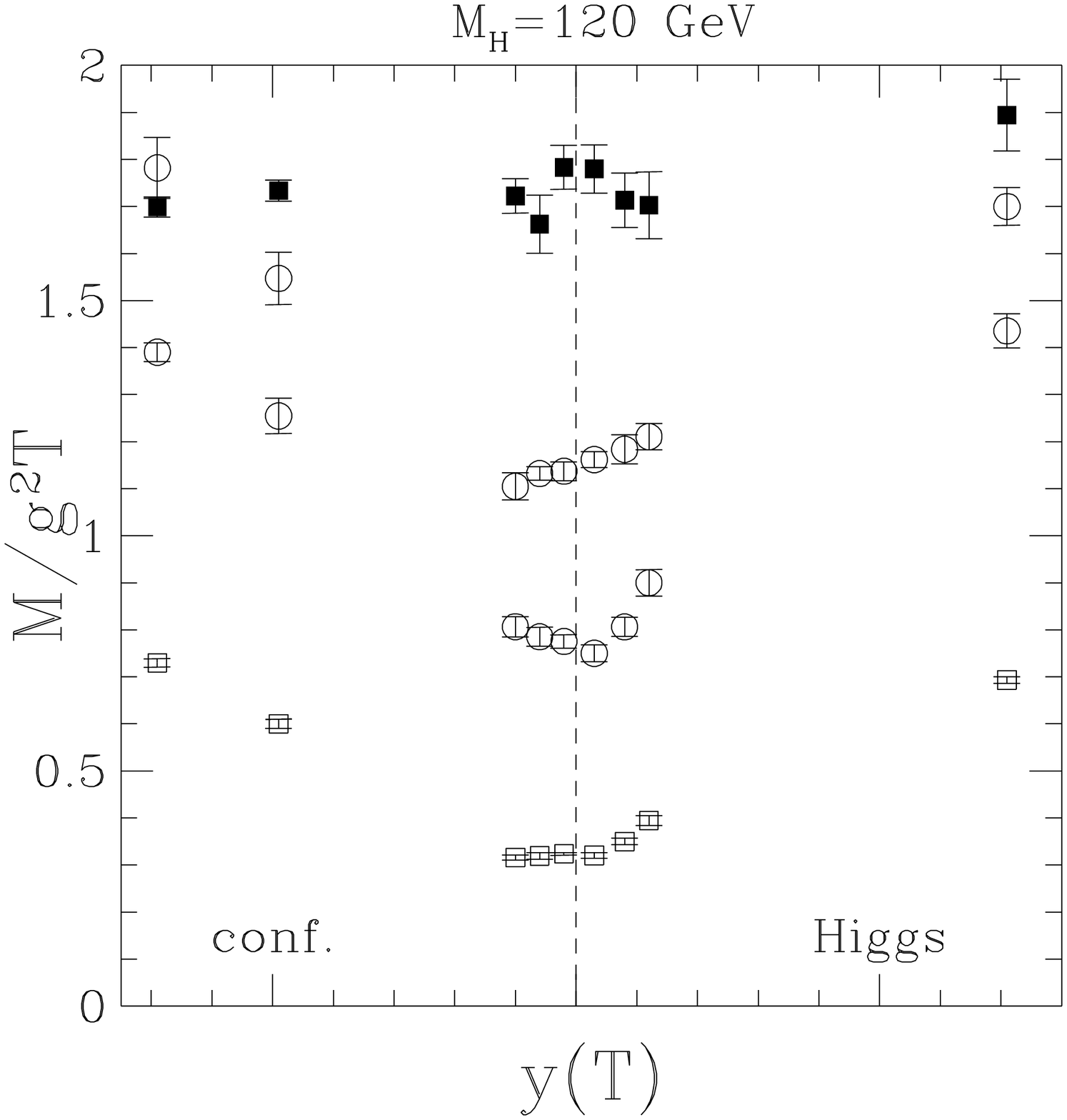}
\leavevmode
\epsfysize=3.6cm
\epsfbox{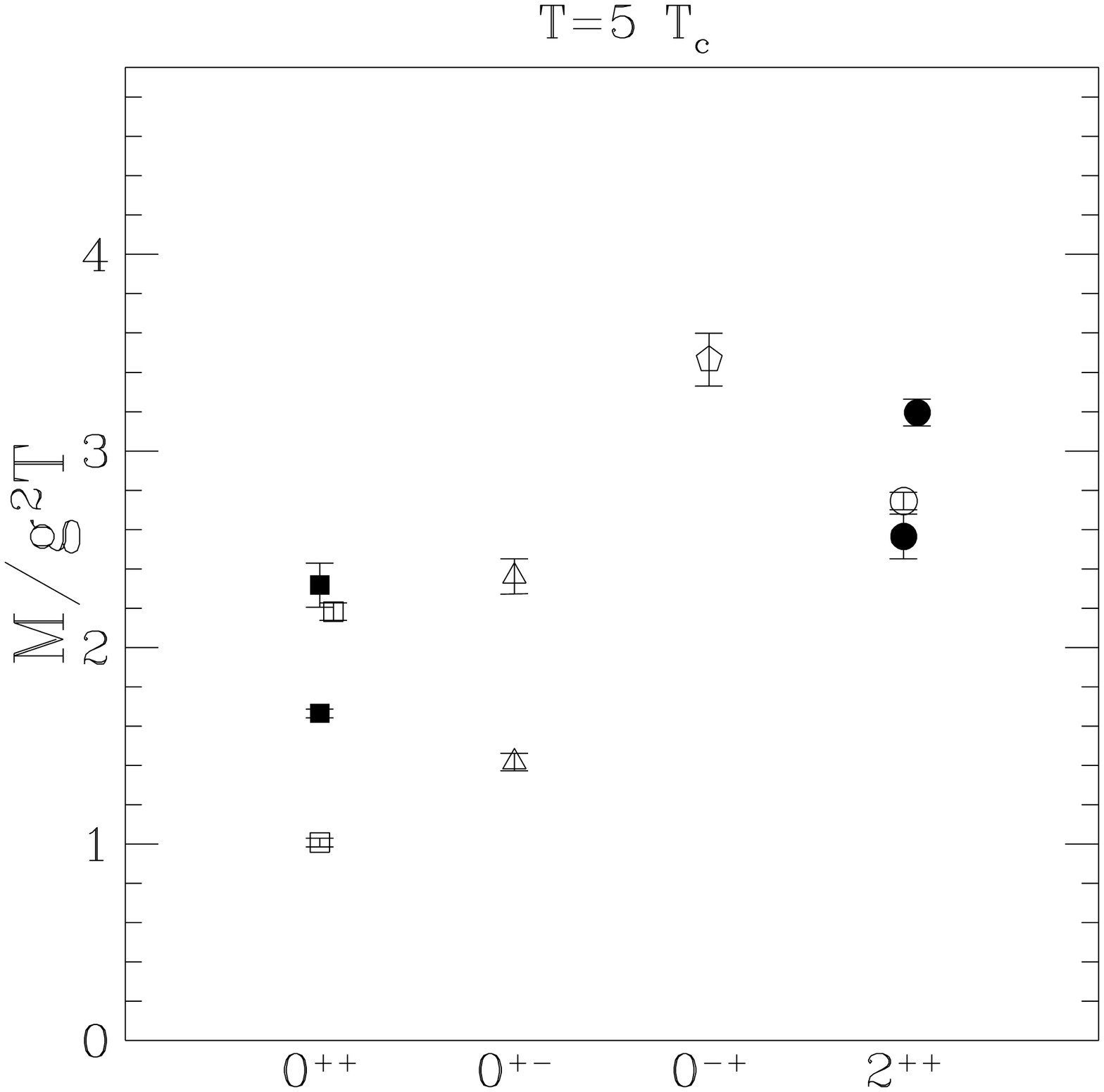}
\caption[]{\label{spec}
Left: Spectrum of static screening masses for 
the SU(2) + fundamental Higgs model for a light Higgs ($x=0.0239$). 
Middle: $0^{++}$ spectrum as function of T for heavy Higgs ($x=0.274$).
Right: Spectrum for SU(2) + adjoint Higgs, corresponding to $T\approx 5 T_c$
($x=0.104$, $y=0.242$) in dim.red. SU(2) QCD.} 
\end{center}
\end{figure}

The physical properties of gauge theories are encoded in
gauge-invariant $n$-point functions. In particular, the mass spectrum
is computed from the exponential fall-off of two-point correlation functions
\be
\lim_{\mid x-y \mid \rightarrow \infty} \langle \phi^{\dag}(x)\phi(y)\rangle
\sim {\rm e}^{-M\mid x-y \mid},
\ee
where $\phi$ generically denotes some gauge-invariant operator
with quantum numbers $J^{PC}$.
The results of lattice calculations \cite{ptw} of the lowest states
of the spectra at various points in the parameter space of our models
are displayed in Fig.~\ref{spec}.

First, consider the electroweak model
for small Higgs mass, 
when confinement and Higgs region are separated by a first 
order phase transition.
In the Higgs region we see the familiar and perturbatively
calculable Higgs and W-boson in the
$0^{++}$ and $1^{--}$ channels, respectively. There is a large gap
to the higher excitations which are scattering states.
In the confinement region, in contrast, there is
a dense spectrum of bound states in all channels, very much
resembling the situation in QCD.
Open symbols denote bound states of scalar fields, whereas
full symbols represent glueballs. 
At a more realistic large Higgs mass, 
the situation in the confinement region (above the phase transition) is repeated,
but the picture in the Higgs region has changed to a similarly dense spectrum.
No phase transition 
separates the two regimes and the mass spectrum can be continuously connected.
Nevertheless, the two regimes have different dynamics, e.g.~no confining
gauge string is formed between static sources in the Higgs regime right of the
dashed line, whereas 
the properties of the glueball spectrum are entirely insensitive to variations of 
the scalar parameters to the left of it. The line marks the center of a rapid
but smooth transition between these regimes.

Next, consider the SU(2) adjoint Higgs model, corresponding to dimensionally
reduced \cite{dimr} SU(2) QCD with $N_f=0$. 
Fig.~\ref{spec} right shows preliminary
results \cite{al} for some low 
lying states at a point corresponding to $T\approx 5T_c$.
In contrast to the fundamental Higgs model one can also define 
gauge-invariant operators odd under charge conjugation. Otherwise the
situation is completely analogous, with again a repetition of the
pure gauge glueball spectrum, denoted by the full symbols, and bound states
of adjoint scalars, and similarly little dependence of the gauge part on the
scalar parameters.

The properties of the pure gauge sector of the two types of Higgs model
in their confinement phases
are compared with those of pure gauge theory \cite{mike1} in Table \ref{comp}.
The masses of the glueballs as well as the string tension in the 
linear part of the potential agree remarkably well between the three models,
demonstrating that the dynamics of the gauge degrees of freedom is almost 
entirely insensitive to the presence of matter fields.
\begin{table}[t]
\begin{center}
\begin{tabular}{|c|c|c|c|}\hline
               & $a\sqrt{\sigma}$ & $am_G$ & $am^*_G$\\ \hline
pure gauge & 0.1622(4) &   0.755(7) & 1.09(2)\\ \hline
fundamental Higgs & 0.1575(5) & 0.74(3)& 1.08(4) \\ \hline
adjoint Higgs  & 0.1560(70) & 0.74(1) & 1.03(5) \\ \hline
\end{tabular}
\end{center}
\caption[]{\label{comp}
Comparison of string tension, the mass of the $0^{++}$ glueball and its first
excitation in lattice units as measured for equal lattice spacing, 
$\beta=4/(ag_3^2)=9$.}
\end{table}

\section{Static potential and screening}
 
Another quantity determining the physical properties
of confining theories is the potential energy of static colour sources,
which is calculated from the exponential decay of large Wilson loops,
\be
V(r)=-\lim_{t\to\infty}
\frac{1}{t}\ln W(r,t).
\ee
As in four dimensions, a string of colour flux connects the sources,
both in fundamental and adjoint representation, leading to a 
potential rising linearly with their separation.
For the fundamental potential in pure gauge theory, 
this linear rise
continues to infinity. If fundamental matter
fields are present as in the Higgs model, the string breaks at
some scale $r_b$, when its energy is large enough to
produce a pair of scalars. 
This results in a saturation of the potential at a constant
value corresponding to the energy of two static-light mesons which are 
formed after string breaking,
$V(r\rightarrow\infty)= const.$
Considering adjoint sources instead,
string breaking occurs already in pure gauge theory, because the
adjoint string can couple to pair-produced gluons.

\begin{figure}[t]
\begin{center}
\vspace*{-3.5cm}
\epsfysize=7cm
\epsfbox{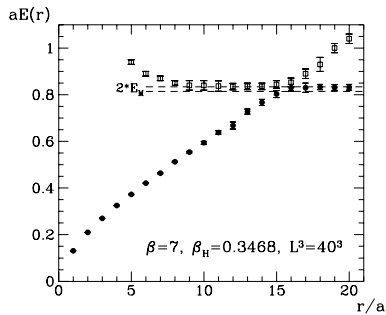}
\caption{\label{break}
Static potential between fundamental sources in the confinement region
of the SU(2) + fundamental Higgs model.}
\vspace*{-0.2cm}
\end{center}
\end{figure}

The string breaking scale $r_b$ is a physical
quantity characterizing
the range of the confining force. Its size depends on the string tension and the
mass of the dynamical particles that have to be produced to break
the string. For the fundamental 
potential in the SU(2) Higgs model it thus depends on the 
bare scalar mass in the Lagrangian. 
In a recent lattice simulation \cite{pw} $r_b$ has been computed for the
same parameter values as the light Higgs confining spectrum 
(cf. Fig.~\ref{spec})
by extracting it from the turnover of the potential
as shown in Fig.~\ref{break}. The continuum extrapolation of
those results gives
$r_b g_3^2\approx 8.5$, $r^{-1}_b\approx 0.12 g_3^2$.
For comparison,
the lightest scalar bound state from Fig.~\ref{spec} is
$m_S=0.839(15)g_3^2$, and the lightest glueball $m_{G}=1.60(4)g_3^2$.
On the other hand, considering the adjoint potential in pure gauge
theory, there is no bare mass in the Lagrangian allowing to tune the mass of the 
constituents, and hence $r_b$ is a purely dynamical quantity of the theory.
In this case the continuum result is \cite{pw} $r_b g_3^2=6.50(94)$, or 
$r_b m_G=10.3\pm 1.5$. In other words, 
the 3d pure gauge theory contains a mass scale $r_b^{-1}$ which 
is by an order of magnitude smaller than the mass of the lightest physical state.

\section{The Debye mass}
 
An important concept in the phenomenology of high temperature
QCD is the static electric screening mass, or the Debye mass $m_D$.
Although its leading order contribution is perturbative,
it couples to the 3d magnetic sector in next-to-leading
order, and hence requires a non-perturbative treatment as 
\nopagebreak
well.
The Debye mass can be expanded as \cite{rebhan}
\be
m_D = m_D^\rmi{LO}+{Ng^2T\over4\pi}\ln{m_D^\rmi{LO}\over
g^2T} + c_N g^2T + {\cal O}(g^3T),
\label{md4d}
\ee
where $m_D^\rmi{LO}=(N/3+N_f/6)^{1/2}gT$
and $N_f$ is the number of flavours.
The logarithmic part of the ${\cal O}(g^2)$ correction
can be extracted perturbatively \cite{rebhan},
but $c_N$ and the higher terms are non-perturbative.
To allow for a lattice determination, a non-perturbative
definition was formulated \cite{ay} employing
the SU(N) adjoint Higgs model as
the dimensionally reduced effective theory.
By integrating out the heavy adoint Higgs
this is further reduced to the pure SU(N) theory in 3d.
The coefficient $c_N$ 
can then be determined from the exponential fall-off
of an adjoint Wilson line $U^{\rm adj}_{ab}(x,y)$
with appropriately chosen adjoint charge operators at the ends, for example
\be
G_{F}(x,y)=
\langle F^a_{ij}(x) U^{\rm adj}_{ab}(x,y)F^b_{kl}(y)\rangle . 
\ee
From its measurement in a lattice simulation
of 3d pure SU(2) one finds the complete non-perturbative
$O(g^2T)$ corrections to the Debye mass with high precision
to be \cite{mo} $c_2=1.06(4)$. Comparing this correction with the 
($N_f=0$) leading term,
one finds $c_2g^2T/m_D^\rmi{LO}=1.3 g$ which is close to one
even for couplings smaller than one.

\section{Summary}
Three-dimensional gauge theories exhibit a rich structure of physical 
mass scales. Although all of them are necessarily $\sim {\cal O}(g_3^2=g^2T)$,
the coefficients vary by more than an order of magnitude. 
For finite temperature field theory this means that 
non-perturbatively the hierarchy of scales in the 
purely magnetic sector is larger than that
between the electric and the magnetic sector, which are not generally well 
separated for realistic couplings.


\section*{References}
\begin{thebibliography}{99}

\bibitem{dimr}
K.Kajantie et al.,
Nucl.Phys.B458 (1996) 90; Nucl.Phys.B503 (1997) 357.

\bibitem{pt}
W.Buchm\"uller, O.Philipsen, Nucl.Phys.B443 (1995) 47;\\
K.Kajantie et al., Phys.Rev.Lett.77 (1996) 2887; \\
A.Hart et al., Phys.Lett.B396 (1997) 217.

\bibitem{ptw}
O.Philipsen et al., Nucl.Phys.B469 (1996) 445; \\
Nucl.Phys.B528 (1998) 379.

\bibitem{mike1}
M.Teper, Phys.Rev.D59 (1999) 014512.

\bibitem{al}
A.Hart, O.Philipsen, work in progress.

\bibitem{pw}
O.Philipsen, H.Wittig, Phys.Rev.Lett.81 (1998) 4056; 
hep-lat/9902003.

\bibitem{rebhan}
A.K.Rebhan, Phys.Rev.D48 (1993) R3967;
Nucl.Phys.B430 (1994) 319.

\bibitem{ay}
P.Arnold, L.Yaffe, Phys.Rev.D52 (1995) 7208.

\bibitem{mo}
M. Laine, O. Philipsen,
Nucl.Phys.B523 (1998) 267.

\end{thebibliography}
\end{document}